\documentclass[iop]{emulateapj}
\usepackage[breaklinks,colorlinks,urlcolor=blue,citecolor=blue,linkcolor=blue]{hyperref}
\usepackage{enumitem}
\usepackage{academicons}
\usepackage{xcolor,graphicx,xspace,color,longtable}
\usepackage{amssymb,amsmath,shadow,bezier,curves,rotating}
\usepackage{graphicx,chngcntr}
\graphicspath{ {./images/}}

\usepackage[toc,titletoc]{appendix}
\usepackage[flushleft]{threeparttable}

\shorttitle{Mass-richness relation and Cosmological parameters of $\mathtt{GalWCat19}$ Clusters}
\shortauthors{Abdullah et al. 2022}

\newcommand {\h}  {$h^{-1}\,$Mpc}
\newcommand {\ks} {km s$^{-1}$}
\newcommand {\hm} {$h^{-1} \  M_{\odot}$}

\newcommand {\sig} {$\sigma_8$}
\newcommand {\om} {$\Omega_\mathrm{m}$}

\begin{document}

\title{Constraining Cosmological Parameters using the Cluster Mass-Richness Relation}

\author{Mohamed H. Abdullah\hyperlink{$^{1}$}{$^{1}$}$^,$\hyperlink{$^{2}$}{$^{2}$}}
\author{Gillian Wilson\hyperlink{$^{3}$}{$^{3}$}}
\author{Anatoly Klypin\hyperlink{$^{4}$}{$^{4}$}$^,$\hyperlink{$^{5}$}{$^{5}$}}
\author{Tomoaki Ishiyama\hyperlink{$^{1}$}{$^{1}$}}

\affiliation{\hypertarget{$^{1}$}{$^{1}$}Institute of Management and Information Technologies, Chiba University, 1-33, Yayoi-cho, Inage-ku, Chiba, 263-8522, Japan}
\affiliation{\hypertarget{$^{2}$}{$^{2}$}Department of Astronomy, National Research Institute of Astronomy and Geophysics, Cairo, 11421, Egypt}\email{melha004@ucr.edu}
\affiliation{\hypertarget{$^{3}$}{$^{3}$}Department of Physics and Astronomy, University of California Riverside, 900 University Avenue, Riverside, CA 92521, USA}
\affiliation{\hypertarget{$^{4}$}{$^{4}$}Astronomy Department, New Mexico State University, Las Cruces, NM 88001}
\affiliation{\hypertarget{$^{5}$}{$^{5}$}Department of Astronomy, University  of Virginia, Charlottesville, VA 22904, USA}

\begin{abstract} 
The cluster mass-richness relation (MRR) is an observationally efficient and potentially powerful cosmological tool for constraining the mean matter density of the universe \om~and the amplitude of fluctuations \sig\ using the cluster abundance technique. We derive the MRR relation using $\mathtt{GalWCat19}$, a publicly available galaxy cluster catalog we created from the Sloan Digital Sky Survey-DR13 spectroscopic dataset. The MRR shows a tail at the low-richness end. Using the Illustris-TNG and mini-Uchuu cosmological numerical simulations, we demonstrate that this tail is caused by systematical uncertainties. We show that, by means of a judicious cut, identified by the use of the Hinge function, it is possible to determine a richness threshold above which the MRR is linear i.e., where cluster mass scales with richness as $\log{M_{200}}= \alpha+\beta \log{N_{200}}$. 
We derive the MRR and show it is consistent with both sets of simulations with a slope of $\beta\approx 1$. We use our MRR to estimate cluster masses from the $\mathtt{GalWCat19}$ catalog which we then use to set constraints on \om~and \sig. Utilizing the all-member MRR, we obtain constraints of \om = $0.31^{+0.04}_{-0.03}$ and \sig= $0.82^{+0.05}_{-0.04}$, and utilizing the red-member MRR, we obtain \om = $0.31^{+0.04}_{-0.03}$ and \sig= $0.81^{+0.05}_{-0.04}$. Our constraints on \om~and \sig~are consistent and very competitive with the \textit{Planck~2018} results.
\end{abstract}

\keywords{ galaxies: clusters: general - cosmology - cosmological parameters}

\section{Introduction}\label{sec:Intro}
In the current picture of structure formation, galaxy clusters arise from rare high peaks of the initial density fluctuation field. These peaks grow in a hierarchical fashion through the dissipationless mechanism of gravitational instability with more massive halos growing via continued accretion and merging of low-mass halos \citep{White91,Kauffmann99,Kauffmann03}. Galaxy clusters are the most massive virialized systems in the universe and are amongst the most effective probes of cosmology. An accurate measurement of the cluster abundance can be used to determine cosmological parameters. The cluster mass function (CMF\footnote{Throughout the paper we use CMF for mass functions derived from observations and HMF for
mass functions computed by theoretical models.}), or the abundance of galaxy clusters, is particularly sensitive to two cosmological parameters in particular,~\om, the matter density of the universe and \sig, the rms mass fluctuation on the scale of 8 \h~at $z=0$ (e.g., \citealp{Wang98,Battye03, Dahle06,Wen10}).

One of the biggest limitations to utilizing the cluster abundance as a cosmological tool is the difficulty of obtaining an accurate estimate of cluster masses because cluster mass is not a directly observable quantity. It can be estimated indirectly using different methods, such as, the virial mass estimator (e.g., \citealp{Binney87}), weak gravitational lensing (e.g., \citealp{Wilson96,Holhjem09}), or application of the Jeans equation of the gas density calculated from the X-ray analysis of a galaxy cluster (e.g., \citealp{Sarazin88}). However, these methods are observationally expensive, require high-quality data sets, and are biased owing to the assumptions that have to be made (e.g., spherical symmetry, hydrostatic equilibrium, and galaxies as tracers of the underlying mass distribution). Fortunately, the cluster mass can still be indirectly inferred from other observables, the so called mass proxies, which scale tightly with cluster mass. Among these mass proxies are X-ray luminosity, temperature, the product of X-ray temperature and gas mass (e.g., \citealp{Pratt09,Vikhlinin09,Mantz16a}), optical luminosity (e.g., \citealp{Yee03}), and the velocity dispersion of member galaxies (e.g., \citealp{Biviano06,Bocquet15}).  In particular, cluster richness (total number of member galaxies with luminosities larger than a certain luminosity threshold) is extensively used in the literature (e.g., \citealp{Pereira18,Abbott20}) as a mass calibration, and commonly referred to as the mass-richness relation (hereafter MRR).

\begin{figure*}\hspace{0.0cm}
    \includegraphics[width=\linewidth]{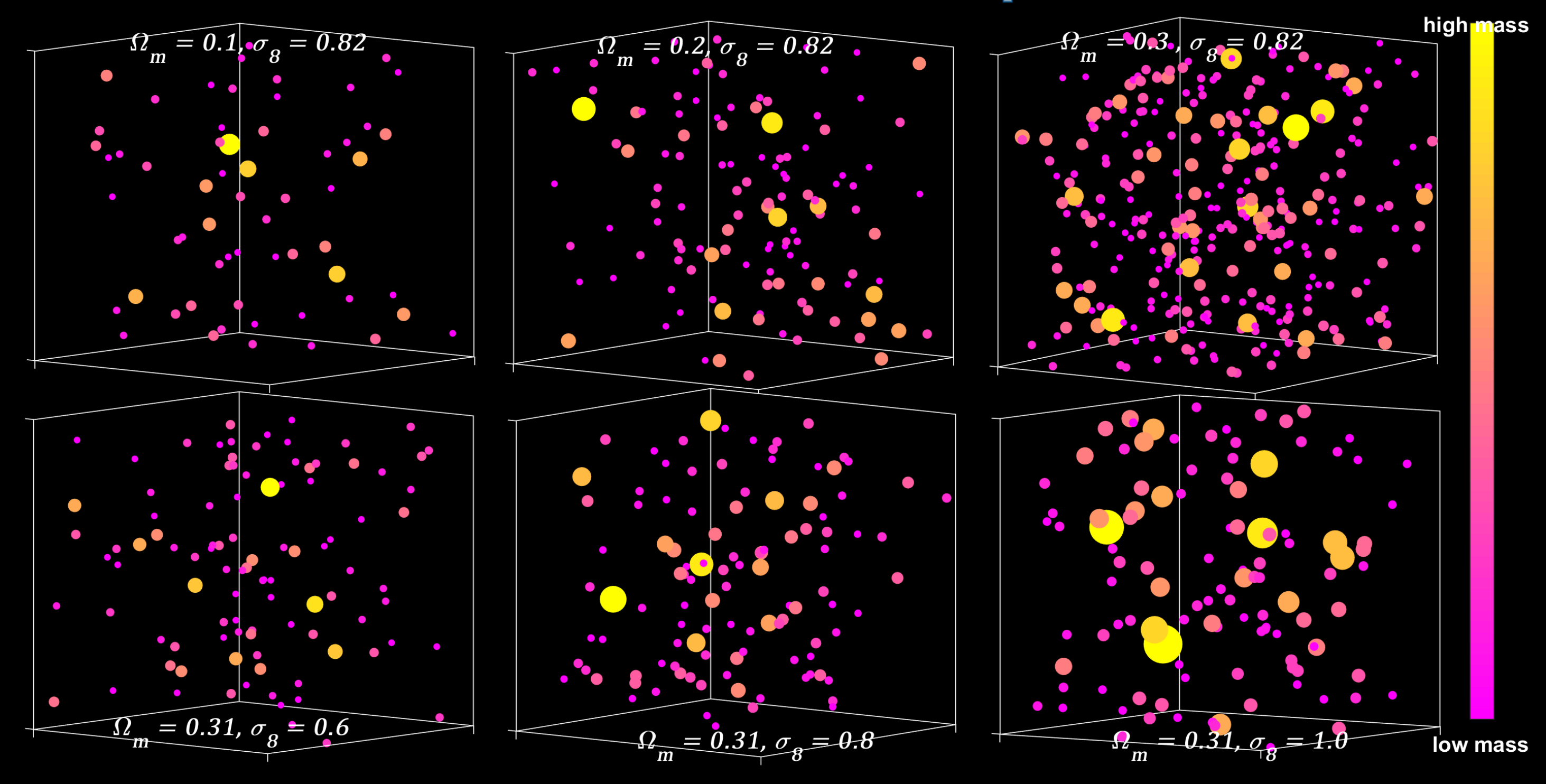} \vspace{0.0cm}
    \caption{Schematic diagram showing the effect on the number of clusters and their masses of varying \om~(upper) and \sig~(lower) independently while holding the other parameter fixed. Each circle represents a cluster with its size (large to small) and color (yellow to magenta) indicating high to low mass. A larger value of \om\ results in an approximately proportionally higher number of clusters of every mass.  A larger value of \sig\ also results in a higher number of clusters of every mass but also changes the  {\textit ratio} of high to low mass clusters, so results in a larger number of high mass  clusters relative to low mass clusters. Note that the clusters are distributed randomly in each box.}
    \label{fig:Fig01}
\end{figure*}

Another challenge arises because of the well-known degeneracy between \om~and \sig\ : the fact that it is possible to obtain a similar cluster abundance from a variety of different combinations of values of \om~and \sig.
Figure \ref{fig:Fig01} illustrates this degeneracy. It shows the effect on the cluster abundance of fixing each of the two parameters in turn, while varying the other. Using the functional form of the halo mass function provided by \citet{Tinker08} we calculate the HMF and then plot the expected number and masses of clusters within a fixed volume. As shown in Figure \ref{fig:Fig01}, increasing \om~while fixing \sig~results in increased numbers of clusters of all masses. Interestingly, increasing \sig~while fixing \om~also results in increased numbers of clusters of all masses. However, increasing \sig~ increases the number of high-mass clusters more dramatically than the number of low-mass clusters. In other words, \sig~is very sensitive to the high-mass end of the HMF. 
Large wide deep-field imaging and spectroscopic galaxy surveys observing over a range of redshift, such as DES \citep{Abbott18b}, DESI \citep{Levi19}, Euclid \citep{Euclid19}, eROSITA \citep{Merloni12}, the LSST that be carried out at the Vera C. Rubin Observatory \citep{LSST09}, the High Latitude Wide Area Survey that will be carried out at the Nancy Grace Roman Telescope \citep{Akeson19}, SPHEREX \citep{Dore2014}, and the Subaru Prime Focus Spectrograph (PFS) \citep{Takada14} will collectively
and simultaneously increase the precision of the constraints on \om~and \sig, break the degeneracy between them, and also constrain additional parameters such as the dark energy $\Omega_\Lambda$, and the equation of state $\omega$  (e.g., \citealp{Rozo10,Weinberg13,Planck18}). 

The MRR is an observationally cheap but potentially powerful cosmological tool. With a deluge of deep, wide-area multi-passband imaging becoming available from a slew of ground and space surveys, there is renewed interest in exploring the potential of the MRR for constraining cosmological parameters. Masses of individual galaxy clusters can be estimated using the MRR relation, and then used to derive the CMF and constrain \om\ and \sig. This cluster abundance technique using cluster masses derived from the MRR ($\mathrm{CA}_\mathrm{MRR}$; see Table \ref{tab:Abb}) has traditionally been applied to photometric galaxy cluster catalogs which lack estimates of individual cluster masses
e.g. \citealp{Rozo10,Costanzi19,Kirby19,Lesci22}.

Cluster catalogs constructed from photometric surveys provide optical richnesses for a very large number of groups and clusters, but may introduce large systematics. This is because distances inferred from photometric redshift estimates are less accurate than those inferred from spectroscopic redshifts. 
This increases the incidence of line-of-sight galaxies in close projection which are falsely-assigned as cluster members. Moreover, cluster catalogs constructed from photometric surveys do not return an estimate of each cluster’s mass. For such photometric samples the cluster mass must be inferred indirectly from the optical richness, which scales tightly with cluster mass. To estimate cluster masses for these photometric samples it is necessary to follow up a subset of clusters with known masses which are calculated individually using  different mass estimators such as weak lensing or X-ray observations. Cluster masses can also be estimated using stacked weak lensing mass profiles by calculating the mean mass of the clusters in a richness bin (e.g., \citealp{Simet17}). Then, the MRR can be calibrated for these samples. As a result of these challenges (as we discuss further below), the MRRs derived from previous studies (e.g., \citealp{Johnston07,Wiesner15,Melchior17}) are in tension with each other, even in the case of studies which use the same type of cluster sample. This is the main reason for the discrepancies or tensions among the derived cosmological parameters (see \citealp{Abdullah20b}).

To better understand how to apply the MRR in the absence of spectroscopy (spectroscopy sparse), we  analyze a photometric cluster catalog for which spectroscopy exists. Specifically, we derive the MRR, the average mass for a cluster that has $N$ galaxies, using the spectroscopic $\mathtt{GalWCat19}$ cluster catalog. Then we pretend that we do not know the true cluster mass and use MRR to find it. This procedure tests the scatter in MRR  and potential systematics.

The $\mathtt{GalWCat19}$ catalog is in the redshift range of $0.01\leq z_{cl} \leq 0.2$ with masses $M_{200} \geq 0.4\times10^{14}$ \hm ~\citep{Abdullah20a}. The advantages of using the $\mathtt{GalWCat19}$ catalog  are as follow. The $\mathtt{GalWCat19}$ was derived from the Sloan Digital Sky Survey-Data Release 13 spectroscopic dataset (hereafter SDSS DR13\footnote{\url{https://www.sdss.org/dr13/}}, \citealp{Albareti17}). The clusters were first identified by looking for the Finger-of-God effect (see, \citealp{Jackson72,Kaiser87,Abdullah13}). The cluster membership was constructed by applying the GalWeight technique which was specifically designed to simultaneously maximize the number of {\it{bona fide}} cluster members while minimizing the number of contaminating interlopers \citep{Abdullah18}. The cluster masses were calculated individually from the dynamics of the member galaxies via the virial theorem (e.g., \citealp{Limber60,Abdullah11}), and corrected for the surface pressure term (e.g., \citealp{The86,Carlberg97}). A main advantage of our approach is that it returns an estimate of the total cluster mass without making any assumptions about the internal complicated physical processes associated with the baryons (gas and stars). 

In the first part of the paper we derive the relationship between the dynamical mass and the optical richness. Note that we derive the MRR relations for: (i)  all members within $R_{200}$\footnote{Throughout the paper we assume $R_{200}$ is the virial radius of the cluster and $M_{200}$ is the virial mass enclosed within $R_{200}$. In practice, the virial radius at which the cluster is in hydrostatic equilibrium cannot be determined. We follow convention and assume that the virial radius $R_{200}$ is the radius within which the average density $\langle\rho(r<R_{200})\rangle = 200\rho_c$, where $\rho_c$ is the critical density of the universe.} (hereafter $\mathrm{MRR_{all}}$) and  (ii) for the red-only members within $R_{200}$ (hereafter $\mathrm{MRR_{red}}$). Notice that red galaxies are identified using red-sequence region in the color-magnitude diagram (e.g., \citealp{Hao09}, see Appendix \ref{sec:appA}). In the second part of the paper, we utilize the MRR to estimate cluster masses and then to derive the cosmological parameters. For this purpose, we assume that the $\mathtt{GalWCat19}$ catalog provides only the richness of each cluster, but we do not use cluster mass given by the catalog. Instead, we apply the derived MRR relations to estimate the cluster masses. We then construct cluster mass function and derive the cosmological parameters of \om~and \sig.

The paper is organized as follows. In \S~\ref{sec:data}, we briefly describe the $\mathtt{GalWCat19}$ cluster catalog, which we use in deriving the MRR. We discuss how we determine the completeness of the catalog as a function of mass and redshift, and how we determine the richness of each cluster. In \S~\ref{sec:richness}, we describe the basic ingredients and methodology of the MRR analysis and our results of the MRR parameters. In \S~\ref{sec:cosmology}, we present our constraint on the cosmological parameters of \om~and \sig~for the $\mathtt{GalWCat19}$ cluster catalog and compare our results with the previous studies. We summarize our conclusions and future work in \S~\ref{sec:conc}. Throughout the paper we adopt $\Lambda$CDM with $\Omega_\mathrm{m}=1-\Omega_\Lambda$, and $H_0=100$ $h$ km s$^{-1}$ Mpc$^{-1}$. Note that throughout the paper we assume $\log{}$ for $\log_{10}$.

\section{Data} \label{sec:data}
\subsection{The $\mathtt{GalWCat19}$ Cluster Catalog} \label{sec:cat}
In this section we summarize how the $\mathtt{GalWCat19}$ cluster catalog was created. Full details can be found in \citet{Abdullah20a}. Using both the photometric and spectroscopic datasets from SDSS-DR13, we extract data for all galaxies that satisfy the following set of criteria: spectroscopic detection, photometric and spectroscopic classification as galaxy (by the automatic pipeline), spectroscopic redshift between 0.001 and 0.2 (with a redshift completeness $> 0.7$, \citealp{Yang07,Tempel14}), r-band magnitude (reddening-corrected) $< 18$, flag SpecObj.zWarning = zero (indicating a well-measured redshift). This results in a catalog containing 704,200 galaxies satisfying all of the criteria.

Galaxy clusters were identified by the well-known Finger-of-God effect (FoG, \citealp{Jackson72,Kaiser87,Abdullah13}). We applied the binary tree algorithm (e.g., \citealp{Serra11}) to accurately determine the cluster center and a phase-space diagram. Galaxy membership for each cluster was assigned by applying the GalWeight technique (developed by our group and presented in \citet{Abdullah18}) to galaxies in the phase-space diagram out to a maximum projected radius of 10~\h~and within a maximum line-of-sight velocity range of $\pm3000$~\ks. In \citet{Abdullah18}, using the Bolshoi simulation \citep{Klypin16}, we showed that GalWeight was $\sim 98\%$ accurate in assigning cluster membership for clusters with mass $M_{200} > 2 \times 10^{14}~h^{-1}M_{\odot}$ and $\sim 85\%$ for clusters with mass $M_{200} > 0.4 \times 10^{14}~h^{-1}M_{\odot}$.

After applying GalWeight to determine cluster membership, the virial mass of each cluster was estimated. This was done by applying the virial theorem under the assumption that the mass distribution follows the galaxy distribution (e.g., \citealp{Giuricin82,Merritt88}). The estimated mass was then corrected for the surface pressure term which, otherwise, would overestimate the true cluster mass (e.g., \citealp{The86,Binney87}). The cluster virial mass was calculated at the viral radius within which the cluster is in hydrostatic equilibrium. The virial radius is approximately equal to the radius within which the density $\rho=\Delta_{200}\rho_c$, where $\rho_c$ is the critical density of the universe and $\Delta_{200} = 200$ (e.g., \citealp{Carlberg97,Klypin16}). \citet{Abdullah20a} showed that the cluster mass estimates returned by the virial theorem performed very favorably, when compared to other commonly utilized mass estimation techniques, which were described and compared in \citealp{Old15}.

\begin{table*}
    \caption{List of abbreviations used in this paper.}
    \label{tab:Abb}
    \scriptsize
    \begin{tabular}{ll}
    \hline
    Abbreviation & Definition\\
    \hline
    MRR & mass-richness relation\\
    $\mathrm{MRR_{all}}$ & mass-richness relation derived for all member galaxies within $R_{200}$ \\
    $\mathrm{MRR_{red}}$ & mass-richness relation derived for red member galaxies within $R_{200}$ \\
    $\alpha$ & normalization of the mass-richness relation\\
    $\beta$ & slope of the mass-richness relation\\
    $\sigma_\mathrm{int}$ & intrinsic scatter in the mass-richness relation\\
    $N_\mathrm{th}$ & richness threshold\\
    $\mathcal{S}_\mathrm{fid}$ & a fiducial subsample of 756 clusters with $\log{M_{200}} \geq 13.9$ [\hm]~and $0.045 \leq z \leq0.125$\\
    $\mathcal{S}\mathrm{all}_{17}$ & a fiducial subsample of clusters with $\log{M_{200}} \geq 13.9$ [\hm], $0.045 \leq z \leq0.125$ and $N_\mathrm{th} = 17$ for all members within $R_{200}$\\
    $\mathcal{S}\mathrm{red}_{13}$ & a fiducial subsample of clusters with $\log{M_{200}} \geq 13.9$ [\hm], $0.045 \leq z \leq0.125$ and $N_\mathrm{th} = 13$ for red members within $R_{200}$\\
    
    $f_\mathrm{x}$& fractional scatter defined as $f_x = (x-x_{fid})/x_{fid}$,and $x$, and $x_\mathrm{fiid}$ are the estimated and fiducial parameters\\
    MCMC&Markov chain Monte Carlo\\
    CMF& cluster mass function\\
    HMF& halo mass function\\
    $\mathrm{CA}_\mathrm{MRR}$ & deriving constraints on \om~and \sig\ using the cluster abundance technique and cluster mass estimates from the mass-richness \\
    & relation\\
    $\mathrm{CA}_\mathrm{dyn}$ & deriving constraints on \om~and \sig\ using the cluster abundance technique and cluster mass estimates from the dynamics \\
    &of member galaxies\\
    $\mathrm{CMB}$ & deriving constraints on \om~and \sig\ using the cosmic microwave background radiation (CMB) technique\\
    \hline
    \end{tabular}
\end{table*}

The $\mathtt{GalWCat19}$ catalog is publicly available from the  website \url{http://cdsarc.u-strasbg.fr/viz-bin/cat/J/ApJS/246/2}. As described in \citet{Abdullah20a}, it consists of two tables, one characterizing the clusters and another characterizing the member galaxies. In creating the $\mathtt{GalWCat19}$ catalog, a $\Lambda$CDM cosmology with $\Omega_\mathrm{m}=1-\Omega_\Lambda$, and $H_0=100$ $h$ km s$^{-1}$ Mpc$^{-1}$ was assumed. The list of clusters has 1800 clusters with redshifts in the range $0.01 < z < 0.2$ and total masses in the range  $(0.4 - 14) \times 10^{14}h^{-1}M_{\odot}$. The cluster table also contains the coordinates of each cluster on the sky (RA, Dec), redshift, number of members, velocity dispersion, and dynamical mass within four overdensities ($\Delta = 500, 200, 100, 5.5$). Note that merging clusters have been removed. The $\mathtt{GalWCat19}$ galaxy member table contains 34,471 members which were identified to lie within the virial radius at which the density is 200 times the critical density of the Universe. The galaxy table contains the coordinates of each member galaxy and the ID of the host cluster.

In the remainder of this paper, we will be primarily focused on analyzing the $\mathtt{GalWCat19}$ cluster table and will utilize the $\mathtt{GalWCat19}$ galaxy member table only to calculate the richness of each cluster (see \S~\ref{sec:richness}). For brevity, therefore, hereafter when we refer to the $\mathtt{GalWCat19}$ catalog, we are referring to the $\mathtt{GalWCat19}$ cluster table. 

\subsection{Completeness}  \label{sec:comp}
\citet{Abdullah20b} showed that $\mathtt{GalWCat19}$ is incomplete in redshift at  $z > 0.085$ but that it was possible to correct for this incompleteness in redshift if each cluster at $z > 0.085$ was weighted by a selection function given by
\begin{equation} \label{eq:SD}
\mathcal{S}_z (D) = 1.07 \exp\left[{-\left(\frac{D}{293.4}\right)^{2.97}}\right],
\end{equation}
\noindent where $D$ is the comoving distance to the cluster, and with the condition that  $\mathcal{S}_z \leq 1$. Thus, the weight which is applied to any given cluster at redshift $z$ is  $\mathcal{W}_{z} = 1/ \mathcal{S}_z(D)$.
Caution is advised in using $\mathcal{S}_z$ to weight those clusters which are at higher redshift in the $\mathtt{GalWCat19}$ catalog. This is because above a redshift threshold, the weight becomes disproportionately large and introduces a high scatter and bias in the CMF toward the highest redshift clusters in the catalog. Thus, in order to avoid these effects we restrict our sample to a maximum redshift of $z =0.125$.

\begin{figure*}\hspace{0.0cm}
    \includegraphics[width=\linewidth]{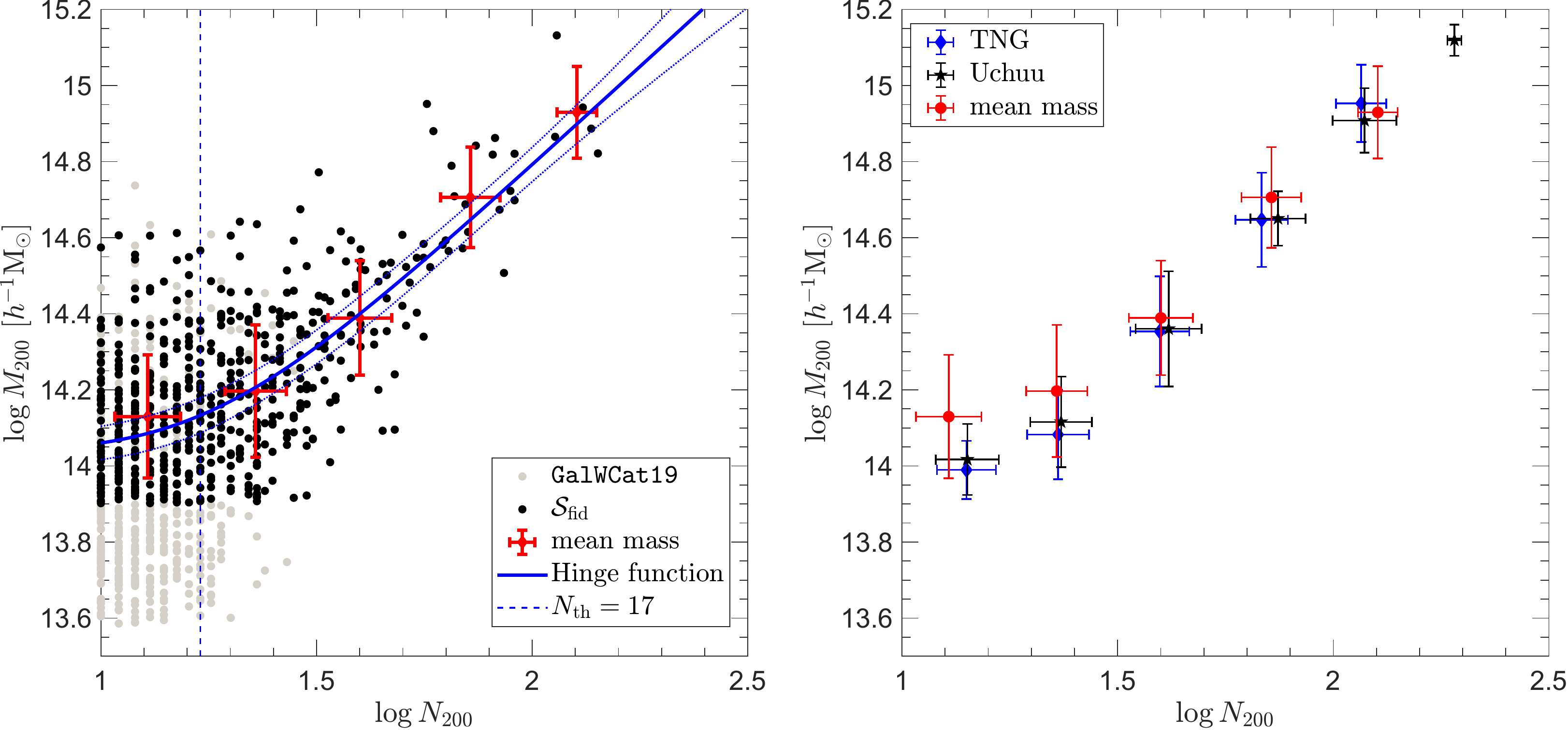} \vspace{-0.25cm}
    \caption{Left: Distribution of clusters from the $\mathtt{GalWCat19}$ catalog in the mass-richness plane. Gray points show all 1800 clusters (with masses of $\log{M_{200}} \geq 13.5$ \hm~and at a redshift $z\leq0.2$ (see \S~\ref{sec:Rich}). Black points show the complete fiducial ($\mathcal{S}_\mathrm{fid}$) subsample of 756 clusters to which the redshift selection function has been applied ($\log{M_{200}} \geq 13.9$ \hm~and $0.045 \leq z \leq0.125$; see \S~\ref{sec:comp}). The mean mass of clusters in $\mathcal{S}_\mathrm{fid}$ as a function of their richness are shown as solid red circles, with error bars indicating 1$\sigma$ Poisson uncertainties. The blue solid line shows the Hinge function for $\mathcal{S}_\mathrm{fid}$ (see~\S~\ref{sec:threshold}), with the dotted blue lines indicating $1\sigma$ uncertainties. The vertical blue dashed line shows the richness threshold at which the slope of the distribution changes (see \S ~\ref{sec:sim} and equation \ref{eq:Hinge}). Right: Red circles and uncertainties are as on left. Also shown are the mean masses of clusters from two simulations (see \ref{sec:sim}) Black points show clusters from mini-Uchuu with a subhalo peak velocity threshold of $v_\mathrm{peak}=130$ \ks~and blue points show clusters from TNG with a galaxy stellar mass threshold of $M_s\geq 5\times 10^9$ \hm. As can be seen clearly from both panels, a flattening (tail) of the MRR occurs at low richness ($\log{N}\lesssim 1.23$). The short tail in the simulations is not  intrinsic. It is partially due to the threshold of simulations as well as Poisson scattering (see \S~ \ref{sec:appD}).}
\label{fig:Fig02}
\end{figure*}

\citet{Abdullah20b} also showed that the 
value of mass at which the catalog is complete depends slightly on the cosmology, but that $\mathtt{GalWCat19}$ is approximately complete for clusters with masses of $\log(M)\geq13.9$ \hm. If we apply the above redshift incompleteness correction and restrict our sample to clusters with redshifts of $0.045 \leq z \leq0.125$ and masses of $\log(M)\geq13.9$ \hm, a total of 756 clusters remain ($\approx 42\%$ of the $\mathtt{GalWCat19}$ sample). We call this fiducial subsample of clusters, $\mathcal{S}_\mathrm{fid}$. In \S~\ref{sec:sys}, we discuss the systematics introduced by adopting these redshift and mass thresholds on our best-fit estimates of \om\ and \sig.

\subsection{Richness} \label{sec:Rich}
In this section we discuss how we calculate the richness of each cluster. For each cluster member, we calculate the absolute magnitude in the $r$-band, $M_\mathrm{r}$, using 
\begin{equation} \label{eq:AbsMag}
M_r - 5 \log {h}= m_r - DM - K(z) - E(z)
\end{equation}
\noindent where $DM(z) = 5 \log{D_L} - 5\log {h}  - 25$ is the distance modulus calculated from the luminosity distance $D_L$, $M_\mathrm{r}$ is the AB apparent magnitude in the $r$-band converted from the SDSS magnitude as $m_{AB}=m_{sdss} +0.010$, $K(z)$ is the $K$-correction, calculated using the latest version of ‘‘Kcorrect’’ (v4), and $E = Q(z-0.1)$  with $Q = -1.62 $ is the evolutionary correction in the $r$-band (see \citealp{Blanton03a,Blanton07} for details). Absolute magnitudes are $K(z)$- and $E(z)$-corrected to redshift $z = 0.1$, which is approximately equal to the mean redshift of the $\mathtt{GalWCat19}$ catalog (Z = 0.089).

We define the richness of each cluster, $N_{200}$, as the total number of members within $R_{200}$ and luminosities $L \geq 0.4 L^\ast$.  $L^\ast$ is the evolved characteristic luminosity of galaxies in the $r$ band defined as $L^\ast(z) = L ^\ast(z=0.1) 10^{Q(z-0.1)}$ and $L^\ast(z=0.1)$ is equivalent to a characteristic absolute magnitude  $M^\ast_r= -20.44$ in the $r$ band \citep{Blanton03a}. This is equivalent to a stellar mass of $\sim 5\times10^9$~\hm (\citealt{Deason19}).

While the GalWeight technique has been shown to be effective at removing foreground/background galaxies  \citep{Abdullah18}, some interlopers which are embedded in the cluster field due to the triple-value problem (see \citealp{Tonry81,Abdullah20a}) still remain in the $\mathtt{GalWCat19}$ catalog and these need to be removed before the MRR can be constructed. As discussed in \citet{Abdullah18}, the percentage of these interlopers is about 11\%.  The total number of galaxies in the cluster, $N_\mathrm{tot}$, is equal to the number of members within $R_{200}$, $N_{200}$, plus the number of interlopers, $N_\mathrm{int}$. In order to remove contamination by these interlopers we calculated the surface density profile of each cluster from its center and fit it with the King equation, defined as
\begin{equation} \label{eq:King}
\Sigma_\mathrm{tot}(R)=\Sigma_c\left(1+\frac{R^2}{r_c^2}\right)^\nu+\Sigma_\mathrm{int},
\end{equation}
\noindent where $\Sigma_c$ and $\Sigma_\mathrm{int}$ are the central and interlopers surface number densities and $r_c$ is a scale radius.  We apply equation~\ref{eq:King} to all members assigned by the GalWeight technique within 6 \h, which is sufficiently large enough to take into account the effect of interlopers. Then, we integrate equation \ref{eq:King} to estimate the contribution of $N_\mathrm{int}$. 

The gray points in the left panel of Figure \ref{fig:Fig02} show the distribution of the 1800 clusters in the $\mathtt{GalWCat19}$ catalog in the mass-richness plane. Visual inspection shows that there is a large scatter in the relationship between mass and richness at the low-richness end. This is due to the large uncertainties introduced when calculating each cluster's mass using a dynamical method when only a handful of member galaxy redshifts are available. The black points show the complete fiducial ($\mathcal{S}_\mathrm{fid}$) subsample of 756 clusters to which the redshift selection function has been applied ($\log{M_{200}} \geq 13.9$ \hm~and $0.045 \leq z \leq0.125$; see \S~\ref{sec:comp}). The mean mass of clusters in $\mathcal{S}_\mathrm{fid}$ as a function of their richness are shown as solid red circles, with error bars indicating 1$\sigma$ Poisson uncertainties. The left panel of Figure \ref{fig:Fig02} shows that when mean mass is plotted as a function of richness, a distinct curve or ``tail" appears to the otherwise linear relationship at the high-richness end. In other words, the MRR has a shallow slope at the low-richness end, while at the high-richness end the relation has a steep slope. Thus, there is a characteristic  richness below which  the trend changes. We call this characteristic richness threshold $N_\mathrm{th}$.   

In the following section, we study two numerical simulations and demonstrate that this tail is not intrinsic. We also demonstrate that it is possible to remove the effect of the tail when deriving the MRR by selecting a subsample of clusters with richness greater than a certain threshold ($N_\mathrm{th}$). This threshold is dependent on the cluster catalog being utilized e.g., \citet{Baxter16} and \citet{Murata19} adopted $N_\mathrm{th}$ = 20 for the redMaPPer cluster catalog \citep{Rykoff16} which utilizes 
red member galaxies, photometrically-selected from the SDSS-DR8 catalog. In \S~ \ref{sec:threshold}, we show how to apply a Hinge function to optimally determine $N_\mathrm{th}$ appropriate for any catalog. Using this method, we derive an optimal threshold of $N_\mathrm{th} = 17~(13)$ for all (red) member galaxies in the $\mathtt{GalWCat19}$ catalog.

\subsection{Simulations} \label{sec:sim}
As discussed in \S~\ref{sec:Rich}, the MRR derived from the $\mathtt{GalWCat19}$ cluster catalog shows a tail at the low-richness end. In this section we analyze two simulations to show that this tail is not intrinsic.
The first simulation is the TNG300-1 simulation from the Illustris TNG300 suite (\citealp{Pillepich18,Nelson19}). 
The TNG300-1 simulation contains $2500^3$ dark matter (DM) particles and the same number of baryonic resolution elements  in a box of comoving length 205 \h. It evolves from redshift $z = 127$ down to $z = 0$ using the AREPO moving-mesh code \citep{Springel10,Weinberger20}, which solves the coupled equations of ideal magneto-hydrodynamics and self-gravity, taking magnetic fields into consideration. It assumes a standard $\Lambda$CDM cosmology, with $\Omega_\Lambda$ = 0.691, $\Omega_\mathrm{m}$ = 0.309, $\Omega_b$ = 0.0486, $h$ = 0.6777, $n_s$ = 0.9667, and $\sigma_8$ = 0.8159 \citep{Planck15}.  The dark matter mass resolution is  $4.0\times 10^7$ \hm~and the baryonic mass resolution is $0.75 \times 10^7$ \hm. The gravitational softening lengths for dark matter and stars in TNG300-1 is  1.0 $h^{-1}$ kpc. In this study,  we use the snapshot \#92 at redshift = 0.08, which most closely matches the mean redshift ($z= 0.089$) of clusters in the $\mathtt{GalWCat19}$ catalog.

The second simulation is the mini-Uchuu simulation from the Uchuu suite of large, high-resolution $N$-body simulations \citep{Ishiyama21} which were done for the Planck2016 cosmology \citep{Planck15}. Mini-Uchuu is a cosmological $N$-body simulation of $2560^3$ particles in a box of co-moving length 400 \h, mass resolution of $3.27 \times 10^8$ \hm, and gravitational softening length of 4.27 $h^{-1}$ kpc. Mini-Uchuu was created using the massively parallel $N$-body TreePM code, \textsc{greem} \citep{Ishiyama09,Ishiyama12}. 
Haloes and subhaloes were identified with \textsc{rockstar} \citep{Behroozi13a} and merger trees constructed with \textsc{consistent trees} \citep{Behroozi13b}. Halo/subhalo catalogs and their merger trees are publicly available through the Skies \& Universes site.\footnote{\url{http://www.skiesanduniverses.org/Simulations/Uchuu/}}
Full details of the Uchuu simulation suite may be found in \citet{Ishiyama21}. Here, we analyze a  snapshot at redshift $z\sim 0.09$. Note that in order to be consistent with how the  $\mathtt{GalWCat19}$ cluster masses are calculated, in the case of both simulations we define the cluster mass $M_{200}$ as the mass enclosed within an overdensity of  $200\rho_c$, where $\rho_c$ is the critical density of the Universe.

The right panel of Figure~\ref{fig:Fig02} shows the MRR obtained from the Illustris-TNG (blue) and mini-Uchuu (black) simulations, as well as from the fiducial $\mathtt{GalWCat19}$ cluster catalog, $\mathcal{S}_\mathrm{fid}$ (red). For the mini-Uchuu simulations we count all subhalos (galaxies) with a threshold of peak velocity $v_\mathrm{peak}= 130$ \ks \ (equivalent to stellar mass of $\sim$ $M_s = 5 \times10^9$ \hm~and luminosity $\sim$ $0.4L_\ast$). For the Illustris-TNG halos (clusters) we determine the number of galaxies with a threshold of stellar mass $M_s = 5 \times10^9$ \hm~within $r_{200}$. Note that for both simulations we select all clusters with masses $\log{M_{200}}\geq 13.9$ \hm. As can be seen from the right panel, there is very good agreements between the $\mathtt{GalWCat19}$ 
MRR and those derived from both Uchuu and TNG at high richness ($N\gtrsim17$, or $\log{N} \geq 1.23$). However, as can clearly also be seen from the right panel, a flattening (tail) of the MRR occurs for both simulations at low richness ($\log{N} \leq 1.23$). In appendix \ref{sec:appD} we show that the length of the tail for both simulations depends on the adopted $v_\mathrm{peak}$ or $M_s$ thresholds. This indicates that the tail is an artifact introduced by the selection applied ($v_\mathrm{peak}$, or equivalently $M_s$), and is, therefore, not real. It is partially due to the threshold of simulations as well as Poisson scattering

Thus, the tail in the MRR obtained from $\mathtt{GalWCat19}$ is not intrinsic and the effect at low richness is due to the Poisson scattering, the systematics of determining the member galaxies and richness, calculating cluster masses of a small members galaxies, and the projection effect. We note that there are very good agreements with MRR obtained from $\mathtt{GalWCat19}$ and both Uchuu and TNG at $N\gtrsim17$. We conclude that it is necessary to apply a cut in richness. In \S~ \ref{sec:threshold} we describe a process for determining the optimal value of richness at which to make the cut.

\subsection{Selecting the richness threshold} \label{sec:threshold}
We wish to determine the optimal richness threshold to apply in order to remove the effect of the tail described in \S~\ref{sec:sim} and Figure \ref{fig:Fig02}. After identifying this value, the MRR can be derived for all clusters with richness larger than this threshold. 
In order to find the richness threshold $N_\mathrm{th}$ at which the slope of MRR changes we use the Hinge function defined as
\begin{equation} \label{eq:Hinge}
\begin{split}
Y = a+b_0(X-X_0)+~~~~~~~~~~~~~~~~~\\ (b-b_0)\delta\log{\left(1+\exp{\frac{X-X_0}{\delta}}\right)},
\end{split}
\end{equation}
\noindent where $Y =\log{M_{200}}$ and $X=\log{N_{200}}$. We use this function only to determine $X_0$, which in our case is the richness threshold $N_\mathrm{th}$.

Applying the Hinge function to the fiducial sample, $\mathcal{S}_\mathrm{fid}$ (\S~\ref{sec:threshold}), we find that the threshold at which the MRR changes its slope is $N_\mathrm{th} = 17$ for all members and $N_\mathrm{th} = 13$ for red members. The blue solid line in the left panel Figure \ref{fig:Fig02} shows the Hinge function for $\mathcal{S}_\mathrm{fid}$ for all members, with the dotted blue lines indicating $1\sigma$ uncertainties. The vertical blue dashed line shows the optimal richness threshold of $N_\mathrm{th} = 17$, derived for all members. In summary, we select all clusters with $\log{M_{200}} \geq 13.9$ [\hm], $0.045 \leq z \leq0.125$, in addition to $N_\mathrm{th} = 17$ for all members and $N_\mathrm{th} = 13$ for red members within $R_{200}$ to derive MRR and cosmological constraints on \om~and \sig. We call these two subsamples $\mathcal{S}\mathrm{all}_{17}$ and $\mathcal{S}\mathrm{red}_{13}$. In \S~\ref{sec:sys} we investigate the systematics of adopting $N_\mathrm{th}$ on our results. 
 
\section{The mass-richness relation}
\label{sec:richness}

In this section we introduce our methodology for fitting the MRR, and then we apply it to the fiducial $\mathtt{GalWCat19}$, $\mathcal{S}\mathrm{all}_{17}$ and $\mathcal{S}\mathrm{all}_{13}$ to derive the best-fit parameters of normalization $\alpha$, slope $\beta$, and intrinsic scatter $\sigma_\mathrm{int}$.

\subsection{Methodology of fitting the mass-richness relation} \label{sec:method} 
The probability distribution of the mass of halos with a fixed richness $N$ is given by a lognormal distribution (e.g., \citealp{Saro15,Simet17,Chiu20}) as
\begin{equation} \label{eq:prob}
\begin{split}
P(\log{M}|N,z)= \frac{1}{\sqrt{2\pi\sigma^2_{\log{M},N}}} \times ~~~~~~~~~~~~~~~~~~~~~\\
\exp{\left[-\frac{\left(\log{M} - \left<\log{M}|N\right>\right)^2}{2\sigma^2_{\log{M},N}}\right]},
\end{split}
\end{equation}

\noindent where the mean mass  $\left<\log{M}|N\right>$ is given as
\begin{equation} \label{eq:rich}
\left<\log{M}|N\right> = \alpha +\beta \log{N},
\end{equation}

\begin{figure}\hspace{-0cm}
    \includegraphics[width=\linewidth]{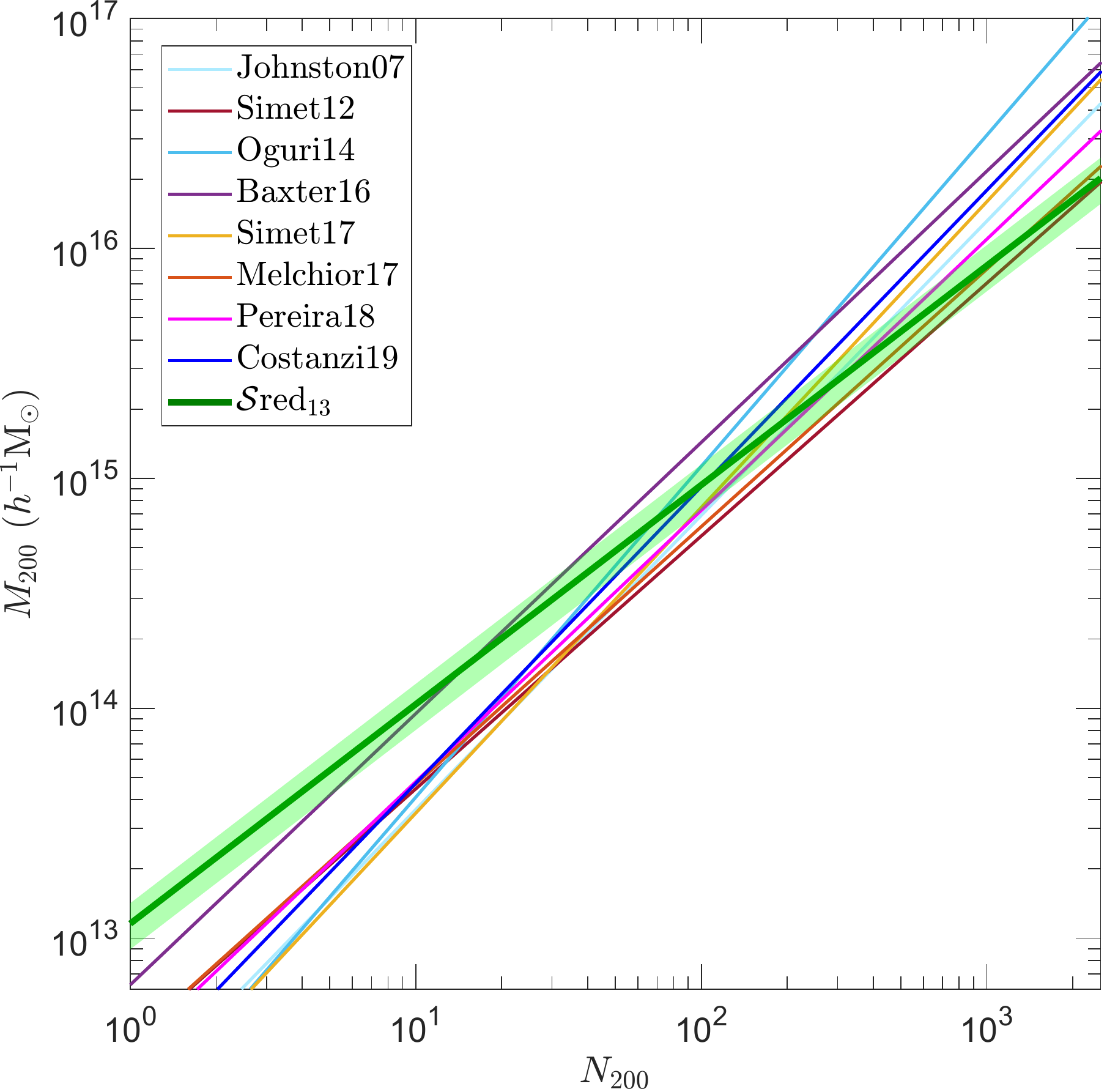} \vspace{-.25cm}
    \caption{Comparison of the best-fit red MRR, $\mathrm{MRR_{red}}$, derived in this work from $\mathcal{S}\mathrm{red}_{13}$ (dark green line with shading indicating $1\sigma$ uncertainty) with results reported in the literature (see legend and Table \ref{tab:MRR}).}
    \label{fig:Fig04}
\end{figure}

In addition, the total variance in the mass $\sigma^2_{\log{M},N}$ at a fixed richness, including contributions of the richness measurement errors $\sqrt{N}$, mass measurement errors $\sigma_{\log{M}}$, and the intrinsic scatter in mass $\sigma_\mathrm{int}$, is modeled by
\begin{equation} \label{eq:var}
\sigma^2_{\log{M},N} = \frac{\beta^2}{N^2}+\sigma^2_{\log{M}}+\sigma^2_{int}
\end{equation}

\noindent where $\alpha$ is the normalization and $\beta$ is the slope of the mass-richness relation. Note that we ignore the evolution term in the MRR relation (Equation \ref{eq:rich}) because our redshift range is very small (0.045-0.0125) and including this team does not affect our results. We estimate the model parameters $\alpha, \beta$, and $\sigma_\mathrm{int}$ with the affine-invariant Markov chain Monte Carlo (hereafter MCMC) sampler of \citet{Goodman10} as implemented in the MATLAB package GWMCMC \footnote{\url{https://github.com/grinsted/gwmcmc}} inspired by the python package $\mathtt{emcee}$ \citep{Foreman13}. 

Using Markov Chain Monte Carlo (MCMC) fitting we now derive best-fit parameters for the MRR within $R_{200}$. For $\mathcal{S}\mathrm{all}_{17}$ we get $\alpha = 12.98\pm0.04$ [\hm], $\beta = 0.96\pm0.03$, and $\sigma_\mathrm{int} = 0.12\pm 0.01$. For $\mathcal{S}\mathrm{red}_{13}$ we obtain $\alpha = 13.08\pm 0.03$ [\hm], $\beta = 0.95\pm $0.02, and $\sigma_\mathrm{int} = 0.11\pm 0.01$.  In \S~\ref{sec:sys}, we discuss the systematics of adopting the redshift, mass, and richness  thresholds on our best-fit estimates of the MRR relation.

\subsection{Comparison to previous results}
\label{sec:comp}
Table~\ref{tab:MRR} compares our best-fit parameters for the red MRR ($\mathrm{MRR_{red}}$)
to those previously published by other groups who analysed different cluster catalogs (see also Figure \ref{fig:Fig04}).
We note that the normalization and slope of our MRR are noticeably different from other analyses. This is because we derived the MRR from the spectroscopic galaxy cluster catalog while other studies used photometric catalogs. As we discussed in the introduction, photometric catalogs have large systematical uncertainties because distances inferred from photometric redshift estimates are less accurate than those inferred from spectroscopic redshifts which increases the systematics of projection effect. Moreover, cluster catalogs constructed from photometric surveys do not return an estimate of each cluster’s mass directly. The masses of these clusters are calibrated via stacking clusters with the same richness and following up a subset of clusters with known masses which are calculated individually using different mass estimators such as weak lensing or X-ray observations (e.g., \citealp{Simet17}).

Also, both the table and the figure demonstrate significant differences in the best-fit parameters amongst the different studies, even for those studies that analyzed the same catalogs e.g.,
\citet{Johnston07} and \citet{Simet12} who utilized the SDSS-MaxBCG catalog of \textcolor{blue}{Koester et al.\ 2007}, and \citet{Baxter16}, \citet{Simet17}, and \citet{Costanzi19} who utilized the SDSS-redMaPPER catalog.

\begin{table*}
    \caption{Comparison of  best-fit parameters for the red mass-richness relation ($\mathrm{MRR_{red}}$) derived here with previously published values using different catalogs.
    All parameters have been calibrated to pivotal richness  $N_\mathrm{piv} = 1$.}
    \label{tab:MRR}
    \scriptsize
    \begin{center}
    \begin{tabular}{ccccc}
    \hline
    Reference&sample-membership& $10^\alpha$ & $\beta$ &redshift\\
    && ($10^{12}$ \hm)    &&\\
    \hline
    \citet{Johnston07}      & photo-SDSS-MaxBCG$^{~(a)}$     & $1.90\pm0.26$  &$1.28\pm0.04$   & $0.10-0.30$ \\
    \citet{Simet12}         & photo-SDSS-MaxBCG              & $3.54\pm0.28$  &$1.10\pm0.12$   & $0.10-0.30$ \\
    \citet{Oguri14}         & photo-SDSS-CAMIRA$^{~(b)}$     & $1.25\pm0.15$  &$1.44\pm0.27$   & $0.10-0.60$ \\
    \citet{Baxter16}        & photo-SDSS-redMaPPer$^{~(c)}$  & $4.84\pm0.96$  &$1.18\pm0.16$   & $0.18-0.33$ \\
    \citet{Simet17}         & photo-SDSS-redMaPPer           & $1.64\pm0.12$  &$1.33\pm0.09$   & $0.10-0.33$ \\
    \citet{Melchior17}      & photo-DES-redMaPPer$^{~(d)}$   & $3.55\pm0.57$  &$1.12\pm0.20$   & $0.20-0.80$ \\
    \citet{Pereira18}       & photo-SDSS-redMaPPer           & $3.17\pm0.12$  &$1.18\pm0.09$   & $0.10-0.33$ \\
    \citet{Costanzi19}      & photo-SDSS-redMaPPer           & $2.42\pm0.17$  &$1.29\pm0.09$   & $0.10-0.30$ \\
    This work ($\mathcal{S}\mathrm{red}_{13}$) & Spec-SDSS-$\mathtt{GalWCat19}$ & $11.4\pm0.16$  &$0.95\pm0.02$ & $0.045-0.125$ \\
    \hline
    \end{tabular}
    \end{center}
    \begin{tablenotes}
    \item
    (a)  A red-sequence cluster finder \citep{Koester07}.
    (b) CAMIRA = Cluster finding algorithm based on Multi-band Identification of red galaxies.
    (c) \citet{Baxter16} used a submsample of SDSS redMaPPer cluster catalog \citep{Rykoff14} which is in the North Galactic Cap (NGC). Here $\log(M_0)$ is calculated at $z = 0.089$ (mean redshift of $\mathtt{GalWCat19}$).
    (d) \citet{Melchior17} used DES (Dark Energy Survey) redMaPPer cluster catalog \citep{Rykoff16}. Here $\log(M_0)$ is calculated at $z = 0.089$.
    (e) \citet{Pereira18} used 230 redMaPPer clusters  obtained from  \citet{Rykoff16} and 136 VT clusters obtained from \citet{Wiesner15} 
    \end{tablenotes}
\end{table*}

The tension in the MRR parameters obtained from different studies can be explained as follows. In deriving the MRR it is necessary to independently calculate the richness and the mass of each cluster. On one hand, estimating cluster richness is complicated. It depends on the cluster-finder method, the projection effect, the completeness of the sample, the definition of the cluster richness, cluster evolution, and the aperture within which the richness is calculated. On the other hand, calculating cluster mass is also complicated. Cluster masses can be calculated by the virial mass estimator, weak gravitational lensing and X-ray observations. However, these methods often return biased results owing to the assumption of hydrostatic equilibrium, projection effect, possible velocity anisotropies in galaxy orbits, the assumption that halo mass follows light (or stellar mass), the presence of substructure and/or nearby structure, the presence of interlopers in the cluster frame(see, e.g., \citealp{Tonry81,The86,Fadda96,Abdullah13,Zhang19}). 

In additional to aforementioned factors, the size of the subsample used for the MRR calibration is usually small (tens of clusters), which introduces large uncertainties in both the slope and the normalization of the MRR relation. Moreover, many cluster catalogs span a large redshift range, so evolution (due to both the evolution of the universe and the physical processes of baryons in clusters) in the scaling relations used to estimate the masses needs to be carefully handled, introducing another source of uncertainty. Other observational systematics that introduce additional uncertainties are photometric redshift errors and cluster miscentering. All of the aforementioned factors can introduce significant uncertainties in the estimates of both the cluster richness and mass and consequently the constraints on MRR parameters (e.g., \citealp{Henry09,Mantz15}) and \om~and \sig~parameters as well (see \S~\ref{sec:Pred} and Figure~\ref{fig:Fig06}).
\section{Implications for cosmological models and constraining \om~and \sig}  \label{sec:cosmology}

In this section we summarize our procedure for constraining \om~and \sig. We start with calculating the halo (cluster) mass function (HMF) from theory, comparing with the CMF we obtain from the MRR relation, and then constraining \om~and \sig. Full details may be found in \citet{Abdullah20b}. 

\subsection{Prediction of Halo Mass Function (HMF)} \label{sec:Pred}
The number of clusters per unit \textcolor{blue}{mass} per unit comoving volume of the universe is given by 
\begin{equation}
\label{eq:hmf}
  \frac{dn}{d\ln M} =  f(\sigma) \frac{\rho_0}{M} \left|\frac{d\ln\sigma}{d\ln M}\right|,
\end{equation}
\noindent where $\rho_0$ is the mean density of the universe, $\sigma$ is the rms mass variance on a scale of radius $R$ that contains a mass $M = 4 \pi \rho_0 R^3/3$ , and $f(\sigma)$ represents the functional form that defines a particular HMF fit. We adopt the functional form of \citet{Tinker08} (hereafter Tinker08) to calculate the HMF and consequently the predicted abundance of clusters. The Tinker08 function is given by 
\begin{equation}\label{eq:T08}
f(\sigma,z) = A\left[\left(\frac{\sigma}{b}\right)^{-a}+1\right] \exp{(-c/\sigma^2)}
\end{equation}  

\noindent where $A = 0.186\left(1+z\right)^{-0.14}$, $a = 1.47\left(1+z\right)^{-0.06}$, $b = 2.57\left(1+z\right)^{-\alpha}$, $c =1.19$, and $\ln{\alpha}(\Delta_{vir}) = \left[75 / \left(\ln{(\Delta_{vir}/75)}\right)\right]^{1.2}$, and $\sigma^2$ is the mass variance defined as
 \begin{equation}\label{eq:s2}
\sigma^2(M,z) =\frac{g(z)}{2\pi} \int P(k)W^2(kR)k^2dk
\end{equation}

\noindent $P(k)$ is the current linear matter power spectrum (at $z=0$) as a function of wavenumber $k$, $W(kR) = 3\left[\sin(kR) - kR\cos(kR)\right])/(kR)^3$ is the Fourier transform of the real-space top-hat window function of radius $R$, and $g(z)=\sigma_8(z)/\sigma_8(0)$ is the growth factor of linear perturbations at scales of 8\h, normalized to unity at $z = 0$. For more details regarding calculation of the HMF we refer the reader to e.g., \citet{Press74,Tinker10b,Behroozi13a,2021ApJ...922...89S}. 

\begin{figure*}\hspace{0cm}
    \includegraphics[width=\linewidth]{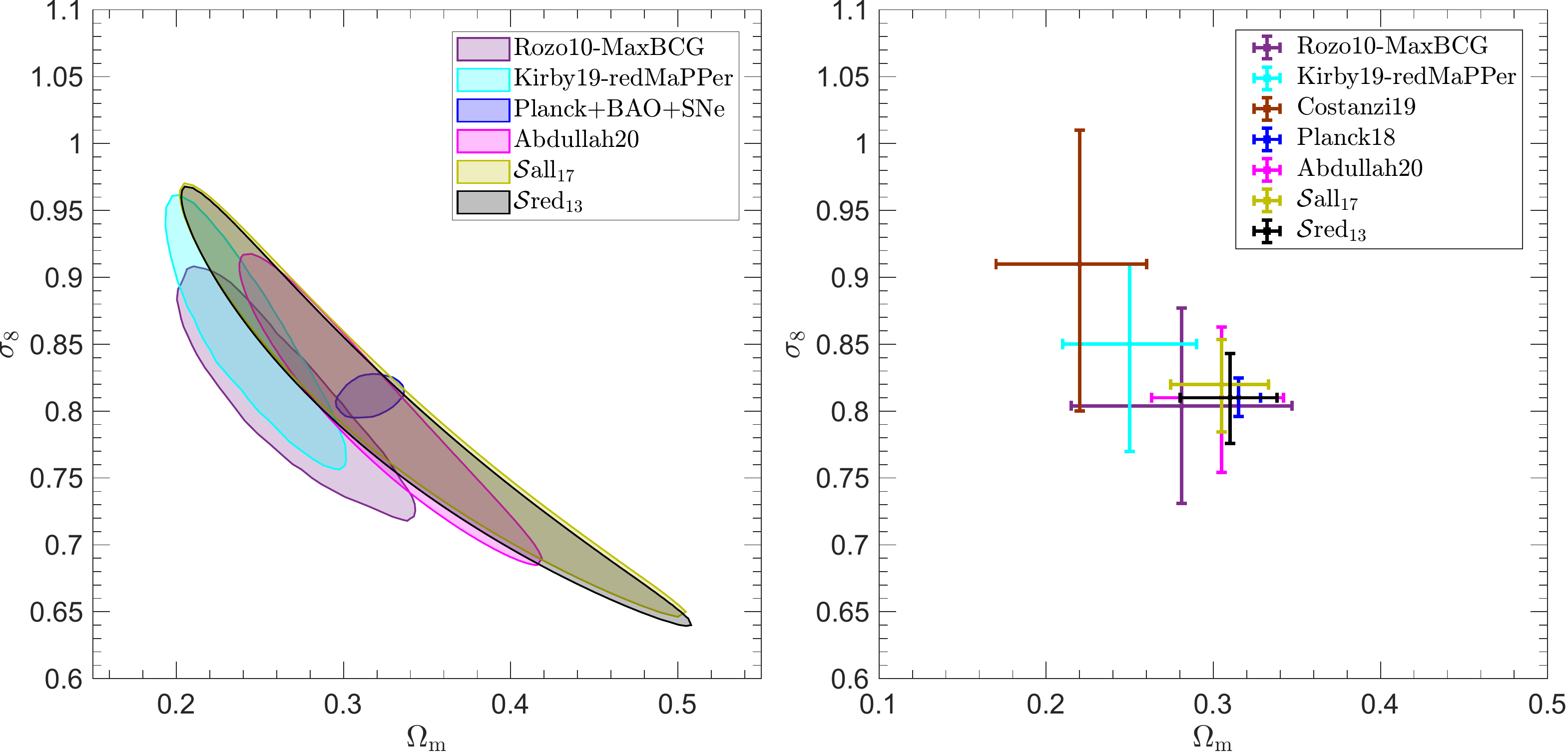} \vspace{0cm}
    \caption{Comparison of constraints obtained on \om~and \sig~in this work with those obtained from previous MRR analyses using different cluster catalogs and richness thresholds. Left: 68\% CLs derived within $R_{200}$ for all members with $N_\mathrm{th}=17$ (yellow, $\mathcal{S}\mathrm{all}_{17}$) and red members with $N_\mathrm{th}=13$ (gray, $\mathcal{S}\mathrm{all}_{13}$). Right: uncertainties on \om~and \sig~estimated from the previous studies of \citet{Rozo10,Costanzi19,Kirby19,Lesci22} (purple, brown,cyan, respectively) which use the cluster abundance technique and cluster mass estimates from the mass-richness relation ($\mathrm{CA}_\mathrm{MRR}$). Also shown are \citet{Abdullah20a} (pink) which uses the cluster abundance technique and cluster mass estimates from the dynamics of member galaxies ($\mathrm{CA}_\mathrm{dyn}$) and \citet{Planck18} (blue) which uses the CMB technique (see Table~\ref{tab:Abb} for the abbreviation).
    }
    \label{fig:Fig06}
\end{figure*}

We calculate the predicted HMF by allowing \om ~to vary between [0.1, 0.6]  and \sig ~between [0.6, 1.2], in both cases in steps of 0.005. We keep the following cosmological parameters fixed: the CMB temperature $T_{cmb}=2.725 K^\circ$, baryonic density $\Omega_b=0.0486$, and spectral index $n = 0.967$ \citep{Planck14}, at redshift $z=0.089$ (the mean redshift of $\mathtt{GalWCat19}$). 

In order to calculate the CMF from observation we begin by deriving MRR relations for both $\mathcal{S}\mathrm{all}_{17}$ and $\mathcal{S}\mathrm{red}_{13}$ subsamples. We then estimate the mass of each cluster knowing its richness and calculate the CMF from observations. The CMF is calculated for clusters with masses $\log{M_{200}} \geq 13.9$ [\hm] and in the redshift range of $0.045 \leq z \leq 0.125$.  \citet{Abdullah20a} showed that, for clusters in the redshift range of $0.045 \leq z \leq 0.125$, the effect of evolution on the HMF is less than 3\%. Note that in this work, rather than utilizing the dynamically-derived estimate of cluster mass in $\mathtt{GalWCat19}$, we re-estimate each cluster mass from its richness using the best-fit MRR and calculate the CMF from those masses.

Finally, in order to determine the best-fit mass function and constrain \om ~and \sig ~we use a standard $\chi^2$ procedure
\begin{equation}
      \chi^2 = \sum_{i=1}^N\left(\frac{\left[y_{o,i}-y_{m,i}\right]^2}{\sigma_i^2} \right),
\end{equation}

\noindent where the likelihood, $\mathcal{L}(y|\sigma_8,\Omega_\mathrm{m})$, of a data $y_o$ given a model  $y_m$ is 
\begin{equation}
      \mathcal{L}(y|\sigma_8,\Omega_\mathrm{m}) \propto \exp{\left(\frac{-\chi^2(y|\sigma_8,\Omega_\mathrm{m})}{2}\right)}
\end{equation}

\noindent Note that $\sigma$ includes the statistical uncertainty of the data plus the intrinsic scatter $\sigma_\mathrm{int}$ obtained from the MRR relation (equation \ref{eq:var}).

\subsection{Constraints on Cosmological Parameters \om~and \sig}

In this section we present the constraints we obtain on \om~and \sig\ using firstly the MRR derived for all members ($\mathcal{S}\mathrm{all}_{17}$) and then for red members ($\mathcal{S}\mathrm{red}_{13}$). 
We also compare the constraints we derive firstly to the cosmological constraints obtained from other groups who fit to MRR relations, and then to the cosmological constraints obtained from the \citet{Planck18} who utilized the CMB technique.

 We derive the cosmological parameters \om~and \sig, as discussed in \S~\ref{sec:Pred}, after estimating each cluster mass in either the full or red subsample using its MRR. 
Utilizing the subsample of clusters with all members, $\mathcal{S}\mathrm{all}_{17}$, we obtain \om = $0.31^{+0.03}_{-0.03}$ and \sig= $0.82^{+0.03}_{-0.04}$. 
Utilizing the subsample of clusters with red members, $\mathcal{S}\mathrm{red}_{13}$, we find \om = $0.31^{+0.03}_{-0.03}$ and \sig= $0.81^{+0.03}_{-0.03}$. Figure~\ref{fig:Fig06} shows our constraints on \om~and \sig~using the subsamples $\mathcal{S}\mathrm{all}_{17}$ and $\mathcal{S}\mathrm{red}_{13}$, as well as previously published constraints from \citet{Rozo10,Costanzi19,Kirby19,Abdullah20a} which use the MRR ($\mathrm{CA}_\mathrm{MRR}$) or dynamical ($\mathrm{CA}_\mathrm{dyn}$) cluster abundance techniques and  \citet{Planck18} which use the CMB technique
(see Table~\ref{tab:Abb} for more details). As shown in the figure, the $68\%$ confidence levels (CLs) obtained from our catalogs $\mathcal{S}\mathrm{all}_{17}$ (yellow) and $\mathcal{S}\mathrm{red}_{13}$ (gray) are very consistent and overlap with each other. Our $68\%$ CLs are also consistent with the $68\%$ CLs obtained by the other groups.
However, despite this overlapping, the right panel of Figure~\ref{fig:Fig06} shows that the constraints on \om~and \sig~obtained from the $\mathrm{CA}_\mathrm{MRR}$ techniques are in tension with each other. This tension comes from the discrepancy between the MRRs derived by the different studies as discussed in \S~\ref{sec:comp}. We end by noting that the constraints we derive on \om~and \sig~\ here from the MRR agree very well both with the constraints derived by the \citet{Planck18} and \citet{Abdullah20a} using the two independent techniques of CMB and cluster dynamics, respectively (although this work and \citet{Abdullah20a} did utilize the same SDSS catalog). In \S~\ref{sec:sys}, we discuss the systematics on the the cosmological constraints introduced by adopting redshift, mass, and richness thresholds.

\begin{figure*}\hspace{0.0cm}
    \includegraphics[width=\linewidth]{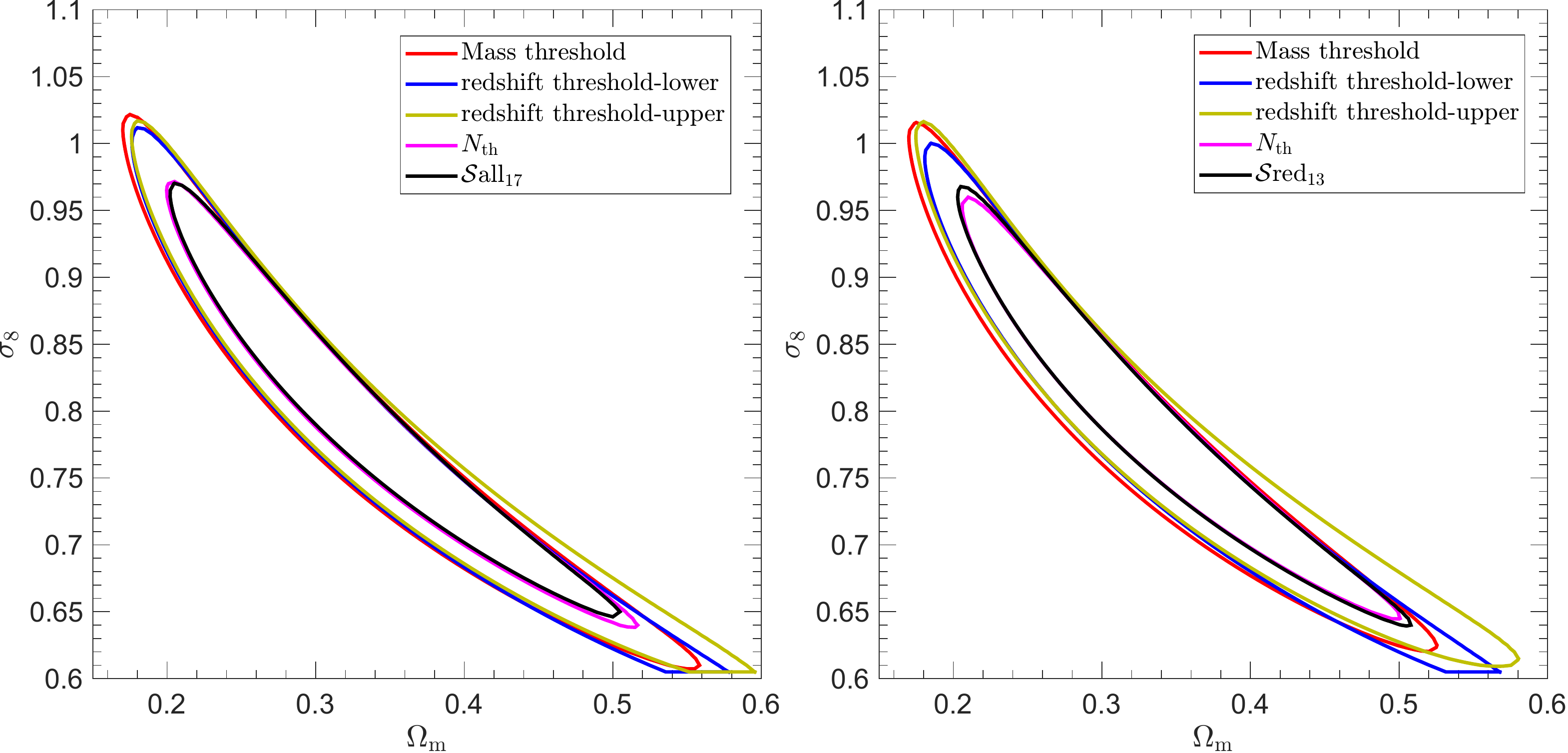} \vspace{-0.25cm}
    \caption{
    Systematical effects of cluster mass, lower and upper redshifts, and richness thresholds on our constraints on the cosmological parameters for the analysis on the fiducial samples of $\mathcal{S}\mathrm{all}_{17}$ for all members within $R_{200}$ (left) and $\mathcal{S}\mathrm{red}_{13}$ for red members within $R_{200}$ (right) (see \S~\ref{sec:sys} for details). The 68\% CLs for our fiducial sample, varying mass threshold $\log M_{200}$ between $13.8$ and $14.0$ [\hm], fixing the upper redshift threshold to 0.125 and varying the lower redshift threshold from 0.01 to 0.07, fixing the lower redshift threshold to 0.045 and varying the upper redshift threshold from 0.11 to 0.15, varying $N_\mathrm{th}$ from 15 to 19 for $\mathcal{S}\mathrm{all}_{17}$, and varying $N_\mathrm{th}$ from 13 to 17 for $\mathcal{S}\mathrm{red}_{13}$.}
    \label{fig:Syst}
\end{figure*}

\begin{table*} \centering
    \caption{Dependence of MRR and cosmological parameters on mass, lower and upper redshifts, and richness  thresholds (\S~\ref{sec:sys}).
    }
    \label{tab:Sys}
    \scriptsize
    \begin{center}
    \begin{tabular}{ccccccc}
    \hline
    &sample&threshold range&$\alpha$& $\beta$&\om&\sig\\
    \hline
    Mass threshold& all &$\log{M_{200,\mathrm{th}}} = [13.8-14.0]$ &$12.97\pm0.10$ &$0.96\pm0.06$ &$0.30\pm0.03$ &$0.82\pm0.02$\\
    
    Mass threshold& red &$\log{M_{200,\mathrm{th}}} = [13.8-14.0]$ 
    &$13.06\pm0.10$ &$0.96\pm0.06$ &$0.30\pm0.02$ &$0.81\pm0.03$\\
    
    redshift lower threshold& all &$z_\mathrm{th}=[0.01-0.07]$ &$12.97\pm0.01$ &$0.96\pm0.01$ &$0.31\pm0.01$ &$0.81\pm0.02$\\
    
    redshift lower threshold& red &$z_\mathrm{th}=[0.01-0.07]$&$13.07\pm0.01$ &$0.95\pm0.01$ &$0.31\pm0.01$ &$0.81\pm0.01$\\
    
    redshift upper threshold& all &$ z_\mathrm{th} = [0.10-0.15]$ &$13.01\pm0.11$ &$0.93\pm0.05$ &$0.33\pm0.05$ &$0.80\pm0.07$\\
    
    redshift upper threshold& red &$ z_\mathrm{th} = [0.10-0.15]$ &$13.13\pm0.13$ &$0.91\pm0.07$ &$0.34\pm0.09$ &$0.80\pm0.08$\\
    
    richness threshold& all &$N_\mathrm{th}=[15-19]$ &$12.98\pm0.10$ &$0.96\pm0.06$ &$0.31\pm0.03$ &$0.82\pm0.04$\\
    
    richness threshold& red &$N_\mathrm{th}=[11-15]$ &$13.10\pm0.10$ &$0.94\pm0.05$ &$0.32\pm0.03$ &$0.81\pm0.03$\\
    \hline
\end{tabular}
\end{center}
\end{table*}

\subsection{Systematics} \label{sec:sys}
In this section we discuss the systematics introduced when deriving the MRR and cosmological parameters due to the adoption of the cluster mass threshold $\log{M_{200}} \geq 13.9$ [\hm], the lower and upper redshift thresholds $0.045 \leq z \leq 0.125$, and the richness thresholds $N_\mathrm{th} = 17$ for $\mathcal{S}\mathrm{all}_{17}$ and $N_\mathrm{th} = 13$ for $\mathcal{S}\mathrm{red}_{13}$.

The first systematic uncertainty comes from the difficulty of accurately determining the mass threshold at which the sample is mass complete. As discussed in Section \S~\ref{sec:comp} the catalog is approximately complete around $\log{M_{200}} \gtrsim 13.9$ [\hm]. We investigate the effect of varying the mass threshold $\log M_{200}$ between $13.8$ and $14.0$ [\hm] on the derived MRR and cosmological parameters from our analysis. For each mass threshold we calculate the $\chi^2$ likelihood and then we obtain the joint 68\% CL of all $\chi$ distributions as shown in Figure~\ref{fig:Syst}. The systematic uncertainties in MRR and cosmological parameters are listed in Table~\ref{tab:Sys}. Both the plot and the table show that the best-fit value of each parameter deviates very slightly from the results of the fiducial samples $\mathcal{S}\mathrm{all}_{17}$ and $\mathcal{S}\mathrm{red}_{13}$.

The second systematic uncertainty comes from the choice of the redshift interval. We first fix the upper redshift threshold to $z=0.125$ and vary the lower redshift threshold from 0.01 to 0.07. Both Figure~\ref{fig:Syst} and Table~\ref{tab:Sys} indicate that varying the lower redshift threshold does not affect our result for the fiducial samples $\mathcal{S}\mathrm{all}_{17}$ and $\mathcal{S}\mathrm{red}_{13}$. It demonstrates that the evolution effect is unremarkable in this small redshift interval (see also \citealp{Abdullah20a}). We then fix the lower redshift threshold to $z=0.045$ and vary the upper redshift threshold from 0.10 to 0.15. The best-fit value of each parameter deviates slightly from the results of the fiducial samples $\mathcal{S}\mathrm{all}_{17}$ and $\mathcal{S}\mathrm{red}_{13}$.

The third systematic uncertainty comes from the choice of the richness threshold $N_\mathrm{th}$. Therefore, we investigate the effect of varying $N_\mathrm{th}$ between $15$ and $19$ for $\mathcal{S}\mathrm{all}_{17}$ and $N_\mathrm{th}$ between $13$ and $17$ for $\mathcal{S}\mathrm{red}_{13}$ on the derived MRR and cosmological parameters. 
Both Figure \ref{fig:Syst} and Table \ref{tab:Sys} show that the best-fit value of each parameter deviates very slightly from the results of the fiducial samples $\mathcal{S}\mathrm{all}_{17}$ and $\mathcal{S}\mathrm{red}_{13}$.

\section{Conclusion} \label{sec:conc}

In this paper, we derive the mass-richness relation (MRR) and constrained the cosmological parameters \om~and \sig~from the $\mathtt{GalWCat19}$ cluster catalog \citep{Abdullah20a}, which was constructed from the SDSS-DR13 spectroscopic data set. The advantages of using $\mathtt{GalWCat19}$ are (i) we were able to identify clusters, assign membership, and determine cluster centers and redshifts with high accuracy from the high-quality SDSS spectroscopic data set; (ii) cluster membership was determined by the GalWeight technique, which has been shown to be $\sim98\%$ accurate in assigning cluster membership (\citealt{Abdullah18}); (iii) cluster masses were calculated individually using the virial theorem and corrected for the surface pressure term; (iv) $\mathtt{GalWCat19}$ is a low-redshift cluster catalog which eliminates the need to make any assumptions about evolution in cluster mass or evolution in cosmological parameters. We select two cluster subsamples $\mathcal{S}\mathrm{all}_{17}$ with richness threshold $N_\mathrm{th} = 17$ for all members and $\mathcal{S}\mathrm{red}_{13}$ with $N_\mathrm{th} = 13$ for red members both within $R_{200}$ and with $\log{M_{200}} \geq 13.9$ [\hm], $0.045 \leq z \leq0.125$. We summarize our findings below:
\begin{enumerate}
\item The MRR shows a tail at low-richness. Using the Illustris-TNG and mini-Uchuu cosmological numerical simulations, we find that this tail is caused by systematical uncertainties. Benchmarking both against Illustris-TNG and mini-Uchuu numerical simulations, we find that, with a judicious choice of richness threshold, cluster richness scales tightly with cluster mass (see Figure \ref{fig:Fig02}). We conclude that the MRR is a powerful and effective tool for estimating cluster mass and for deriving cosmological constrains  from cluster catalogs.

\item Using MCMC fitting (\S~\ref{sec:method}) we derive the best-fit parameters of the MRR within $R_{200}$ (\S~\ref{sec:richness}). For $\mathcal{S}\mathrm{all}_{17}$ we find $\alpha = 12.98\pm0.04$ [\hm], $\beta = 0.96\pm0.03$, and $\sigma_\mathrm{int} = 0.12\pm 0.01$.For $\mathcal{S}\mathrm{red}_{13}$ we obtain $\alpha = 13.08\pm 0.03$ [\hm], $\beta=0.95\pm0.02$, and $\sigma_\mathrm{int} = 0.11\pm 0.01$.

\item The slope of our MRR is consistent with both the Illustris-TNG and mini-Uchuu numerical simulations while MRRs derived from photometric catalogs return a steeper slope. This is likely because we derived our MRR from a spectroscopic galaxy cluster catalog while the other works used photometric catalogs. Photometric 
redshifts are known to be less accurate than those determined from spectroscopic redshifts. This increases the incidence of line-of-sight galaxies in close projection which are falsely-assigned as cluster members.
\item We estimate a mass for each cluster in  the $\mathtt{GalWCat19}$ catalog using the MRR relation and then derive constraints on the cosmological parameters \om~and \sig~using the cluster abundance technique (\S~\ref{sec:Pred} and \S~\ref{sec:method}). Using the MRR determined from all members, we obtain \om = $0.31^{+0.03}_{-0.03}$ and \sig= $0.82^{+0.03}_{-0.04}$, and for red members, $\mathcal{S}\mathrm{red}_{13}$ we obtain \om = $0.31^{+0.03}_{-0.03}$ and \sig= $0.81^{+0.03}_{-0.03}$. We compare our results to \citet{Rozo10,Costanzi19,Kirby19,Lesci22} who also used the cluster abundance technique and mass-richness relation ($\mathrm{CA}_\mathrm{MRR}$). We also compare our results to \citet{Abdullah20a} who used the cluster abundance technique and  galaxy dynamics technique ($\mathrm{CA}_\mathrm{dyn}$) and with \citet{Planck18} who used the CMB technique (see Table~\ref{tab:Abb} for the abbreviation). We find that the constraints we derive from $\mathcal{S}\mathrm{all}_{17}$ and $\mathcal{S}\mathrm{red}_{13}$ are consistent and overlap with each other. Reassuringly, the constraints we obtain on \om~and \sig~ are consistent with those obtained both by \citet{Planck18} and \citet{Abdullah20a} using independent techniques.
\item We discuss the systematics of adopting mass, redshift, and richness thresholds when deriving our MRR and cosmological parameters (\S~\ref{sec:sys}). We find that the best-fit value of each parameter deviates only slightly from those of the two fiducial samples of $\mathcal{S}\mathrm{all}_{17}$ and $\mathcal{S}\mathrm{red}_{13}$ (Figure \ref{fig:Fig06} and Table \ref{tab:Sys}).
\end{enumerate}

In future work we aim to (i) investigate the stellar mass
and luminosity function of member galaxies; (ii) study the connection between stellar mass (or luminosity) and cluster mass; (iii) investigate the environmental effects on the properties of member galaxies such as size, color, and star formation rate; and (iv) study the correlation function of galaxy clusters and the signature of baryonic acoustic oscillations to constrain cosmological parameters utilizing both the low-redshift spectroscopic $\mathtt{GalWCat19}$ cluster catalog and the recent public data releases of the Uchuu-SDSS galaxy lightcones \citep{Paez22}, the Uchuu-$\nu^2$GC semi-analytic galaxy/AGN \citep{Oogi2022}, and the Uchuu-UniverseMachine galaxy catalogues \citep{Aung22}. We also look forward to apply the Fog-GalWeight (see \citealp{Abdullah18,Abdullah20b}) toolkit to the ongoing and upcoming high-redshift spectroscopic surveys such as DESI \citep{Levi19}, Euclid \citep{Euclid19}, SPHEREX \citep{Dore2014}, and Subaru Prime Focus Spectrograph (PFS) \citep{Takada14}.

\begin{acknowledgments}
We gratefully acknowledge support from the Japan Society for the Promotion of Science through JSPS KAKENHI Grant Number JP21F51024. We gratefully acknowledge support from the National Science Foundation through grant AST-2205189, and from HST program numbers GO-15294 and GO-16300. Support for program numbers GO-15294 and GO-16300 was provided by NASA through grants from the Space Telescope Science Institute, which is operated by the Association of Universities for Research in Astronomy, Incorporated, under NASA contract NAS5-26555. T.I. has been supported by IAAR Research Support Program in Chiba University Japan, MEXT/JSPS KAKENHI (Grant Number JP19KK0344 and JP21H01122), MEXT as ``Program for Promoting Researches on the Supercomputer Fugaku'' (JPMXP1020200109), and JICFuS. We thank Instituto de Astrof\'isica de Andaluc\'ia (IAA-CSIC), Centro de Supercomputaci\'on de Galicia (CESGA) and the Spanish academic and research network (RedIRIS) in Spain for hosting Uchuu Data Release one in the Skies \& Universes site for cosmological simulations. The Uchuu simulations were carried out on Aterui II supercomputer at Center for Computational Astrophysics, CfCA, of National Astronomical Observatory of Japan, and the K computer at the RIKEN Advanced Institute for Computational Science. The Uchuu DR1 effort has made use of the skun@IAA\_RedIRIS and skun6@IAA computer facilities managed by the IAA-CSIC in Spain (MICINN EU-Feder grant EQC2018-004366-P).
\end{acknowledgments}

\setcounter{figure}{0}
\setcounter{table}{0}
\setcounter{equation}{0}
\setcounter{section}{0}

\makeatletter 
\renewcommand{\thesection}{A\@arabic\c@section}
\renewcommand{\thefigure}{A\@arabic\c@figure}
\renewcommand{\thetable}{A\@arabic\c@table}
\makeatother

\begin{figure*}\hspace{0cm}
\includegraphics[width=1\linewidth]{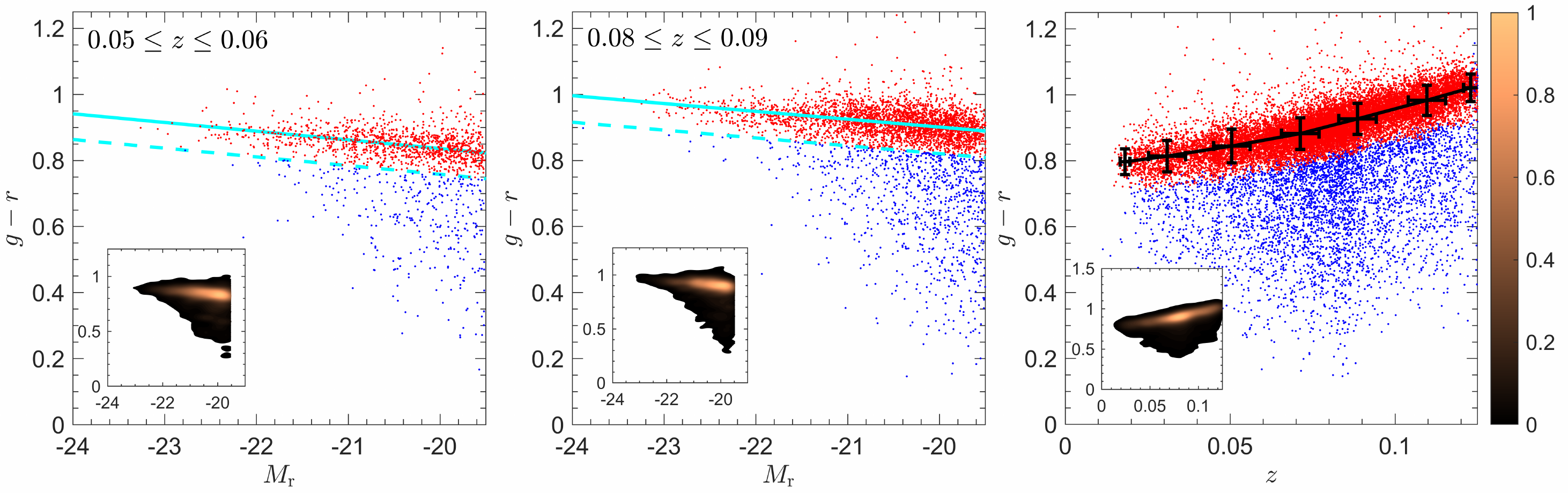} \vspace{-.25cm}
\caption{Separating red and blue members of the $\mathtt{GalWCat19}$ clusters using the color-magnitude diagram (CMD). The left and middle panels show the CMD for cluster members in two different redshift bins. The solid cyan line in each panel shows the red-sequence ridgeline (highest probability) and the dashed line is the line that separates red and blue members at the $1\sigma$ CL from the ridgeline. The right panel shows the evolution of the ridgeline as a function redshift. The black line shows the mean color of the red members as a function of redshift with $1\sigma$ uncertainties. The nested panels show the probability distribution functions for galaxies in the CMD.}
\label{fig:RedBlue}
\end{figure*}

\section{Separating red and blue galaxies} \label{sec:appA}
In this section we describe our procedure to identify the red-sequence galaxies of $\mathtt{GalWCat19}$ members. It is well known that clusters are dominated by elliptical, red E/S0 galaxies which occupy a narrow region in color-magnitude diagram (CMD) known as the E/S0 ridgeline or red-sequence (e.g., \citealp{Bower92,Gladders00}). The location and the slope of this ridgeline in color decreases smoothly with increasing redshift \citep{Koester07}. The $\mathtt{GalWCat19}$ catalog contains clusters in the redshift range $0.01 \leq z \leq 0.2$. In order to identify red members, we sort the clusters into 14 equally-sized redshift bins, each approximately spanning $\Delta z= 0.01$. Then, for each redshift bin, we plot the members in the rest-frame $\mathrm{g-r~vs~M}_r$ CMD to identify the ridgeline. The ridgeline is defined to be the line with maximum likelihood (or pdf) in color for the distribution of galaxies in the CMD (see the first two nested panels in Figure~\ref{fig:RedBlue}). This ridgeline can be located using the two-dimensional adaptive kernel method (2DAKM, see e.g \citealp{Pisani96}) or the Gaussian Mixture Model for red-sequence galaxy identification \citep{Hao09}. In this paper we apply the 2DAKM to identify the location of the ridgeline. We then locate the line with probability of $1\sigma$ CL below the ridgeline. We define red galaxies as those with rest-frame color above the $1\sigma$ line.

Figure \ref{fig:RedBlue} shows the CMD of $\mathtt{GalWCat19}$ galaxies at two redshift bins $0.05 \leq z \leq 0.06$ and $0.08 \leq z \leq 0.09$. The solid cyan line in each panel represents the red-sequence ridgeline and the dashed line is the line that separates the red and blue galaxies with probability of $1 \sigma$ CL from the ridgeline. In the right panel of Figure \ref{fig:RedBlue} we plot the color-redshift relation. The plot demonstrates that the slope of the ridgeline varies with redshift.

\section{Is the MRR tail intrinsic?} \label{sec:appD}
\begin{figure*}\hspace{0.0cm}
    \includegraphics[width=\linewidth]{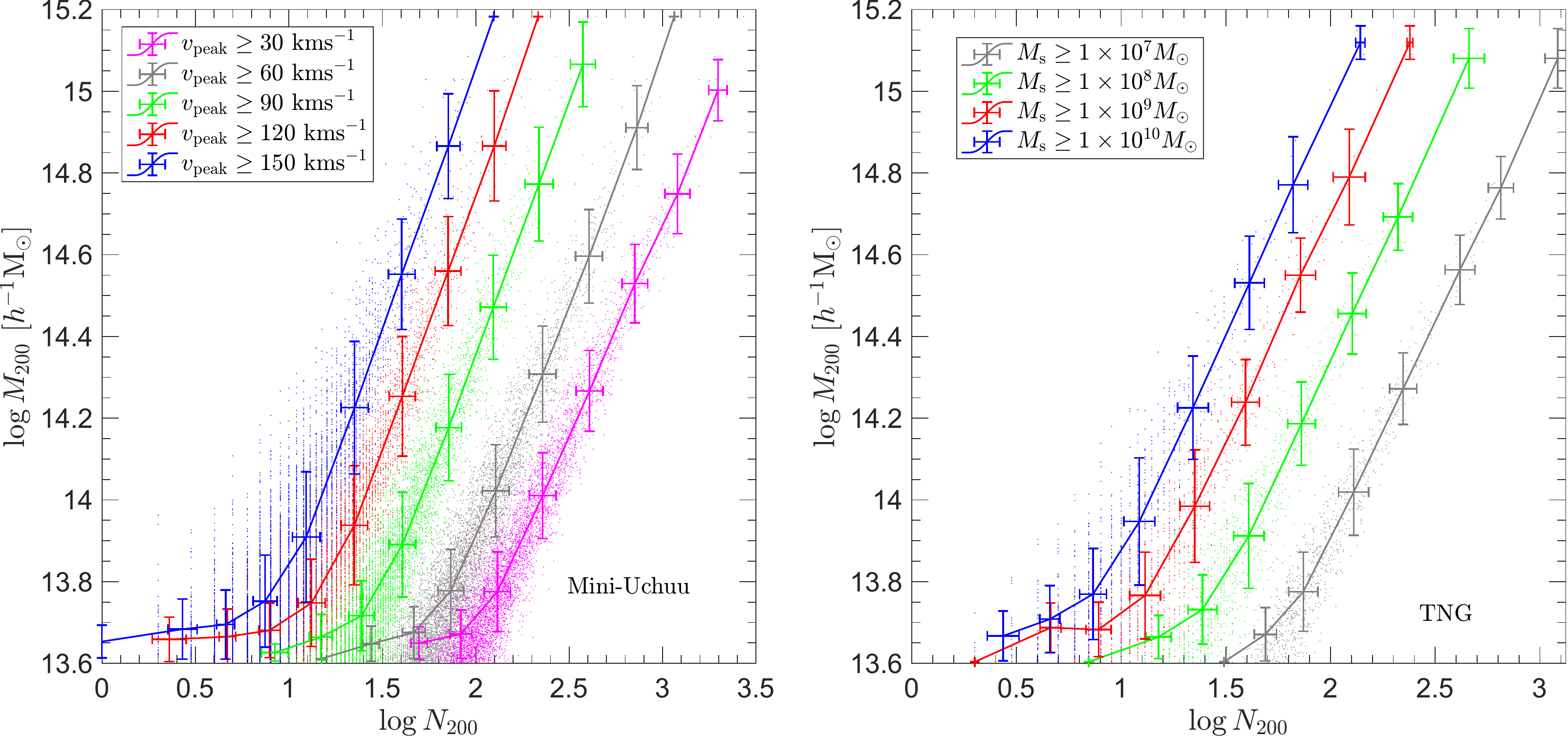} \vspace{-.25cm}
    \caption{Mass-richness relation (MRR) for simulations.  The mean mass at some richness bins is plotted for Mini-Uchuu (left) for five $v_\mathrm{peak}$ thresholds for subhalos and TNG (right) for four $M_s$ thresholds for galaxies within $r_{200}$ as shown in each legend. Error bars represent Poisson noise. The figure shows that MRR introduces a tail at low-richness end. The length of the tail increase with increases the thresholds of $v_\mathrm{peak}$ and $M_s$. This indicates that the tail is dependent on the selection of the threshold and it is not intrinsic.
    }
    \label{fig:Fig03Sim}
\end{figure*}

In Figure \ref{fig:Fig03Sim}, we plot the MRR relation 
 for the mini-Uchuu and Illustris-TNG simulations. For the mini-Uchuu simulations we plot MRR for five $v_\mathrm{peak}$ thresholds for subhalos and for the Illustris-TNG halos we present MRR for four $M_s$ thresholds for galaxies both within $r_{200}$. The figure shows a short tail at low-richness end. The tail increases with increasing the thresholds of $v_\mathrm{peak}$ for the Mini-Uchuu and $M_s$ for TNG of the selected sample from each simulation. We conclude that this tail is not intrinsic and the effect at low richness is partially due to the threshold of simulations as well as Poisson scattering.

\bibliography{ref1}

\end{document}